\documentclass[a4paper,11pt]{article}
\pdfoutput=1
\usepackage{jinstpub}
\usepackage{graphicx,wrapfig}
\usepackage{breqn}
\usepackage{multirow}
\usepackage{lineno}

\begin{document}
\setpagewiselinenumbers

\title{Improved Modeling of $\beta$ Electronic Recoils in Liquid Xenon Using LUX Calibration Data}

\collaboration{The LUX Collaboration}

\author[a,b]{D.S.~Akerib, }
\affiliation[a]{SLAC National Accelerator Laboratory, 2575 Sand Hill Road, Menlo Park, CA 94205, USA} \affiliation[b]{Kavli Institute for Particle Astrophysics and Cosmology, Stanford University, 452 Lomita Mall, Stanford, CA 94309, USA} 
\author[c]{S.~Alsum, } \affiliation[c]{University of Wisconsin-Madison, Department of Physics, 1150 University Ave., Madison, WI 53706, USA}  
\author[d]{H.M.~Ara\'{u}jo, } \affiliation[d]{Imperial College London, High Energy Physics, Blackett Laboratory, London SW7 2BZ, United Kingdom}  
\author[e]{X.~Bai, } \affiliation[e]{South Dakota School of Mines and Technology, 501 East St Joseph St., Rapid City, SD 57701, USA}  
\author[f]{J.~Balajthy, } \affiliation[f]{University of California Davis, Department of Physics, One Shields Ave., Davis, CA 95616, USA}  
\author[g]{A.~Baxter, } \affiliation[g]{University of Liverpool, Department of Physics, Liverpool L69 7ZE, UK}  
\author[h]{E.P.~Bernard, } \affiliation[h]{University of California Berkeley, Department of Physics, Berkeley, CA 94720, USA}  
\author[i]{A.~Bernstein, } \affiliation[i]{Lawrence Livermore National Laboratory, 7000 East Ave., Livermore, CA 94551, USA}  
\author[a,b]{T.P.~Biesiadzinski, } 
\author[h,j,k]{E.M.~Boulton, } \affiliation[j]{Lawrence Berkeley National Laboratory, 1 Cyclotron Rd., Berkeley, CA 94720, USA} \affiliation[k]{Yale University, Department of Physics, 217 Prospect St., New Haven, CT 06511, USA}
\author[g]{B.~Boxer, } \
\author[l]{P.~Br\'as, } \affiliation[l]{LIP-Coimbra, Department of Physics, University of Coimbra, Rua Larga, 3004-516 Coimbra, Portugal}  
\author[g]{S.~Burdin, } 
\author[m,n]{D.~Byram, } \affiliation[m]{University of South Dakota, Department of Physics, 414E Clark St., Vermillion, SD 57069, USA} \affiliation[n]{South Dakota Science and Technology Authority, Sanford Underground Research Facility, Lead, SD 57754, USA} 

\author[o]{M.C.~Carmona-Benitez, } \affiliation[o]{Pennsylvania State University, Department of Physics, 104 Davey Lab, University Park, PA  16802-6300, USA}  
\author[p]{C.~Chan, } \affiliation[p]{Brown University, Department of Physics, 182 Hope St., Providence, RI 02912, USA}  

\author[f]{J.E.~Cutter, } 
\author[o]{L.~de\,Viveiros, } 
\author[q]{E.~Druszkiewicz, } \affiliation[q]{University of Rochester, Department of Physics and Astronomy, Rochester, NY 14627, USA}  
\author[a,b]{A.~Fan, }
\author[j,p]{S.~Fiorucci, } 
\author[p]{R.J.~Gaitskell, } 
\author[r]{C.~Ghag, } \affiliation[r]{Department of Physics and Astronomy, University College London, Gower Street, London WC1E 6BT, United Kingdom}  

\author[j]{M.G.D.~Gilchriese, } 
\author[g]{C.~Gwilliam, }
\author[s]{C.R.~Hall, } \affiliation[s]{University of Maryland, Department of Physics, College Park, MD 20742, USA}  

\author[t]{S.J.~Haselschwardt, } \affiliation[t]{University of California Santa Barbara, Department of Physics, Santa Barbara, CA 93106, USA}  
\author[j,u]{S.A.~Hertel, } \affiliation[u]{University of Massachusetts, Amherst Center for Fundamental Interactions and Department of Physics, Amherst, MA 01003-9337 USA} 
\author[h]{D.P.~Hogan, } 
\author[h,n]{M.~Horn, } 
\author[p]{D.Q.~Huang, } 
\author[a,b]{C.M.~Ignarra, }
\author[h]{R.G.~Jacobsen, } 
\author[r]{O.~Jahangir, } 
\author[a,b]{W.~Ji, } 
\author[h,j]{K.~Kamdin, } 
\author[i]{K.~Kazkaz, }  
\author[q]{D.~Khaitan, } 
\author[v]{E.V.~Korolkova, } \affiliation[v]{University of Sheffield, Department of Physics and Astronomy, Sheffield, S3 7RH, United Kingdom}  
\author[j]{S.~Kravitz, }   
\author[v]{V.A.~Kudryavtsev, }
\author[w]{E.~Leason, } \affiliation[w]{SUPA, School of Physics and Astronomy, University of Edinburgh, Edinburgh EH9 3FD, United Kingdom}  
\author[f,i]{B.G.~Lenardo, } 
\author[j]{K.T.~Lesko, }  
\author[p]{J.~Liao, }  
\author[h]{J.~Lin, }
\author[l]{A.~Lindote, }
\author[l]{M.I.~Lopes, }  
\author[f]{A.~Manalaysay, }
\author[c,x]{R.L.~Mannino, } \affiliation[x]{Texas A \& M University, Department of Physics, College Station, TX 77843, USA}  
\author[d]{N.~Marangou, }
\author[w]{M.F.~Marzioni, }
\author[h,j]{D.N.~McKinsey, }
\author[m]{D.-M.~Mei, } 
\author[q]{M.~Moongweluwan, }  
\author[f]{J.A.~Morad, }  
\author[w]{A.St.J.~Murphy, }
\author[v]{A.~Naylor, }
\author[t]{C.~Nehrkorn, } 
\author[t]{H.N.~Nelson, }  
\author[l]{F.~Neves, }
\author[w]{A.~Nilima, }
\author[h,j]{K.C.~Oliver-Mallory, }
\author[c]{K.J.~Palladino, }
\author[h,j]{E.K.~Pease, }  
\author[h,j]{Q.~Riffard, } 
\author[y,1]{G.R.C.~Rischbieter, \note{Corresponding author.}} \affiliation[y]{University at Albany, State University of New York, Department of Physics, 1400 Washington Ave., Albany, NY 12222, USA}  
\author[p]{C.~Rhyne, } 
\author[v]{P.~Rossiter, }
\author[r,t]{S.~Shaw, } 
\author[a,b]{T.A.~Shutt, } 
\author[l]{C.~Silva, }
\author[t]{M.~Solmaz, }  
\author[l]{V.N.~Solovov, }
\author[j]{P.~Sorensen, }
\author[d]{T.J.~Sumner, }
\author[y]{M.~Szydagis, } 
\author[n]{D.J.~Taylor ,}
\author[d]{R.~Taylor, }
\author[p]{W.C.~Taylor, }
\author[k]{B.P.~Tennyson, }
\author[x]{P.A.~Terman, }  
\author[s]{D.R.~Tiedt, }
\author[z]{W.H.~To} \affiliation[z]{California State University Stanislaus, Department of Physics, 1 University Circle, Turlock, CA 95382, USA}  
\author[f]{M.~Tripathi, }
\author[h,j,k]{L.~Tvrznikova, }
\author[r]{U.~Utku, }
\author[f]{S.~Uvarov, }  
\author[d]{A.~Vacheret, }
\author[h]{V.~Velan, }
\author[x]{R.C.~Webb, }
\author[x]{J.T.~White, } 
\author[a,b]{T.J.~Whitis, }
\author[j]{M.S.~Witherell, } 
\author[q]{F.L.H.~Wolfs, } 
\author[o]{D.~Woodward, }
\author[i]{J.~Xu, }
\author[m]{C.~Zhang}




\emailAdd{grischbieter@albany.edu}

\abstract{ We report here methods and techniques for creating an improved model that reproduces the scintillation and ionization response of a dual-phase liquid and gaseous xenon time projection chamber. Starting with the recent release of the Noble Element Simulation Technique (NEST v2.0), electronic recoil data from the $\beta$ decays of ${}^3$H and ${}^{14}$C in the Large Underground Xenon (LUX) detector were used to tune the model, in addition to external data sets that allow for extrapolation beyond the LUX data-taking conditions. This paper also presents techniques used for modeling complicated temporal and spatial detector pathologies that can adversely affect data using a simplified model framework. The methods outlined in this report show an example of the robust applications possible with NEST v2.0 framework and how it can be modified to produce a final, detector-specific, electronic recoil model. This example provides the final model for LUX and detector parameters that will used in the new analysis package, the LUX Legacy Analysis Monte Carlo Application (LLAMA), for accurate reproduction of the LUX data. As accurate background reproduction is crucial for the success of rare-event searches, such as dark matter direct detection experiments, the techniques outlined here can be used in other single-phase and dual-phase xenon detectors to assist with accurate ER background reproduction.}

\begin{NoHyper}
\maketitle
\end{NoHyper}


\section{Introduction}
\label{sec:intro}

The Large Underground Xenon (LUX) Experiment was a dual-phase time projection chamber (TPC) equipped with 370 kg of liquid xenon (LXe), of which 250 kg was an active target. The LUX experiment took place in the Davis Cavern at the 4850' level of the Sanford Underground Research Facility (SURF) in Lead, South Dakota with the primary scientific purpose of detecting WIMP dark matter \citep{lux}. As a dual-phase TPC, LUX collected the prompt scintillation signal (S1) with two arrays each consisting of 61 photo-multiplier tubes (PMTs) located at the top and bottom of the detector. The ionized electrons were extracted with an applied electric field into a layer of gaseous xenon, where the electrons were accelerated to produce a secondary light signal (S2).  The location of the S2 signal in the top PMT array provides the position of the event in the horizontal plane, $x$ and $y$, while the time between the S1 and S2 pulses provides the depth of the interaction, $z$.

While being hosted in the Davis cavern, LUX was used to perform two WIMP searches. The first science run (WS2013) spanned from April to August 2013 \cite{luxFirst, lux2},  while the second run (WS2014-16) started in September 2014 and ended in May 2016 \cite{luxRun4}. The total exposure for the combined runs was $3.35\times10^4$ kg$\cdot$days. LUX was decommissioned in September 2016.

Throughout the WS2014-16 science run, LUX's drift field was significantly non-uniform due to an accumulation of excess charge on the inner detector walls. The distribution of excess charge changed over the course of the WIMP search, creating a temporal and position-dependent drift field. Sophisticated modeling was performed to create accurate three-dimensional maps of LUX's electric fields and is described in detail in Ref.~\cite{fieldModeling}.

 In addition, the scintillation and electroluminescence gain factors, $g_1$ and $g_2$\footnote[2]{$g_1$ and $g_2$ are defined with respect to the expectation values, $\langle S1\rangle = g_1n_\gamma$ and $\langle S2\rangle = g_2n_e$, where $n_\gamma$ and $n_e$ are the numbers of photons and electrons that escape the interaction site after an energy deposition, respectively.}, changed throughout the course of WS2014-16 \cite{luxRun4}. WS2013 did not experience these detector effects in a significant manner.

In existing and upcoming large target LXe dark matter experiments, the most burdensome backgrounds are those from $\beta$ decays of Rn daughters \cite{xenon100_1year, pandaX, TDR, LZsens}, underscoring the importance of measuring and reproducing the detector response to electronic recoils. In addition, there exist several dark matter candidate models where the dominant interaction with Standard Model matter occurs via electronic recoils~\cite{axionLUX,mirrorDM}. To characterize the detector response from $\beta$ decays, LUX underwent multiple calibration campaigns measuring the light and charge yields from incident $\beta$ particles, including those from ${}^3$H and ${}^{14}$C sources which have energy spectra extending out to 18.6 and 156 keV, respectively. As $\beta$-Xe interactions occur via electronic recoils (ER), these calibrations were used to characterize the detector's response to ER-producing background radiation. There were multiple ${}^3$H calibrations that took place periodically throughout LUX's stay in the Davis campus, while a single ${}^{14}$C calibration took place after WS2014-16 in August 2016 immediately before LUX was decommissioned. Details of the source injections system and processes can be found in Refs.~\cite{tritium, jonC14}.

With the recent release of the second version of the Noble Element Simulation Technique (NEST v2.0) \cite{nestv2}, the ER calibration responses were compared to the new NEST model. This model is based on empirical fits to data sets from virtually all existing photon and electron yield data taken with Xe targets, including the LUX WS2013 ${}^3$H ER calibration data. NEST then provides efficient calculation of the observable detector response after a recoil event. 

The NEST v2.0 framework was utilized to create the new LUX analysis package, the LUX Legacy Analysis Monte Carlo Application (LLAMA). LLAMA's main purpose is to reproduce the detector response after an energy deposition with minimal deviation between the simulation results and actual data. While NEST v2.0 only allows for the simulation of a temporally static detector, LLAMA allows for interpolation of detector parameters as a function of time. Equipped with the detailed three-dimensional electric field maps, LLAMA can use the NEST models to reproduce all LUX data despite the significant spatial and temporal detector effects observed after WS2013. LLAMA also serves as an example of how to expand upon the NEST v2.0 framework in order to simulated a temporally dynamic detector. 

Since the empirical models in NEST v2.0 were created by fitting to world data of light and charge yields in LXe, it is understandable to expect that NEST would slightly deviate from the data of a single experiment, especially for data that were not included in the fits. As the WS2014-16 and ${}^{14}$C acquisition data sets were not included during the development of NEST v2.0, the model required additional tuning and optimization to consistently reproduce all LUX data. The purpose of this paper is to develop techniques and methodologies to minimize deviations when modeling ER backgrounds in a LXe detector. More generally, this work explores the limits of what is possible for the detailed understanding of the ER response of future large two-phase xenon TPCs. Therefore, we will describe in detail the process of optimizing NEST v2.0's physics models for reproducing LUX data using LUX ER bands in S1c-log(S2c/S1c) space, where S1c and S2c are the \textit{corrected} S1 and S2 signals that take into account possible position-dependence of g$_1$ and g$_2$, while preserving the structure of the NEST v2.0 simulation package. In addition, results using temporal interpolation and the complete position-dependent electric field maps with LLAMA will be shown, displaying an expansion of the NEST v2.0 framework that can accurately account for dynamic detector parameters and data-taking conditions.

\section{Optimizing the Mean Yields Model}

Given the energy of the incident particle the magnitude of the drift field at the interaction point, and the target Xe density, NEST v2.0 calculates the expected total number of particles generated during the interaction, $n_q$, and then it calculates the total electron yield \cite{nestv2}. The sum of electrons and photons produced in an ER event is given by $n_q = n_e + n_{ph} = E/W$, where $E$ is electronic recoil energy, and $W$ is the average work function for scintillation and ionization in LXe and is approximately 13.7 eV~\cite{DahlThesis}. By calculating the charge yields separately, the number of produced photons is simply the difference between total quanta and number of electrons. After the mean yields are calculated, fluctuations about the mean are calculated separately and applied to the result. 

The NEST v2.0 $\beta$ ER charge yield equation is a sum of two sigmoidal functions. To create the LUX yields model for use in LLAMA, and following work reported in Ref.~\cite{JonThesis}, the NEST v2.0 double-sigmoid was reformulated as \begin{dmath} Qy(E, \mathcal{E}) = m_1 + \frac{ m_2 - m_1 }{ ( 1 + (E/m_3)^{m_4})^{m_9}} + m_5 + \frac{ 0 - m_5 }{ ( 1 + (E/m_7)^{m_8})^{m_{10}}}, \end{dmath}  where $Qy$ is the mean electron yield per unit energy, $n_e/E$. The individual $m_i$ are free parameters that can depend on the applied field strength, $\mathcal{E}$, and are tuned so that the first sigmoid models low energy charge yields and the second sigmoid controls the behavior at higher energies. Note that $m_6$ is explicitly set to zero; because $m_6$ controls the low-energy asymptote of the high-energy sigmoid, it has degenerate effects with $m_1$, which controls the high-energy asymptote of the low-energy sigmoid. This double-sigmoid approach allows for the reproduction of older yields models based on first-principles approximations: the Thomas-Imel Box model (TIB) for low energy ER and Doke's formulation of Birk's Law (DB) for high-energy particle tracks \cite{TIB, DB}. The TIB model approximates the recoil as a point-like scatter and assumes spherical symmetry of the produced particle cascade. At higher energies, the DB formulation assumes cylindrical symmetry of the particle cascade about a high-energy particle track. The medium-energy regime between these two idealized approximations is significantly harder to model \cite{nest1}, but the NEST sum of sigmoids allows for a smooth transition in this stitching region. 

Using data from the 
WS2013 and WS2014-16 ${}^3$H calibrations, the deviation between the simulated and actual S1c-log(S2c/S1c) band means was minimized. However, due to the significant temporal and spatial dependencies of the detector's gains and drift field that were observed in WS2014-16, simplifications of the detector geometry were required. Because of the temporal dependencies, the WS2014-16 data was split into four unequal date bins; the ${}^{14}$C calibration that occurred after WS2014-16 is considered to be a part of a fifth separate date bin. In addition, NEST's framework does not allow for straightforward incorporation of the complicated field maps for WS2014-16, so each date bin was divided into four horizontal slices of the detector's fiducial volume -- each corresponding to a 65 $\mu$s window of drift time. The drift time is defined as the time between the beginning of the S1 and S2 pulses and represents the depth of the interaction. Explicitly, from top to bottom of the detector, these four drift windows are: $40 < t < 105$, $105 < t < 170$, $170 < t < 235$, and $235 < t < 300$, where $t$ is the drift time measured in $\mu$s. To begin simulating the S1c-log(S2c/S1c) bands with NEST, each of the date bins was provided a mean g$_1$ and g$_2$ associated with the data taken in the relevant time period. The field maps provide field magnitudes for each three-dimensional spatial coordinate in the detector. By making a histogram of field magnitudes for each cubic millimeter volume element inside a volume slice, it is straightforward to extract the mean field and the standard deviation within a given volume. For each bin of drift times, NEST randomly selects a field value for each event from a uniform distribution with mean and width related to the relevant volume's mean field and standard deviation.

The tuning of the individual $m_i$ reported in Ref.~\cite{JonThesis} provided a robust ER yields model. To preserve this, only one of the $m_i$ was used as a free parameter while creating the LUX yields model and considering only energies and fields relevant to LUX data. As charge yields from ${}^3$H $\beta$ ER fall into the TIB regime and bleed into the medium-energy stitching region, $m_1$ was chosen as the free parameter for minimizing the deviation between the model and the data, as it directly affects the shape of the charge yields in the stitching region. Explicitly, $m_1$ was allowed to float to minimize the test statistic: \begin{dmath}  
\Delta_\mu = |\sum(\mu_{_{NEST}} - \mu_{_{data}})| + \sqrt{\sum (\mu_{_{NEST}} - \mu_{_{data}})^2}\end{dmath}, where $\mu$ represents the mean value of log(S2c/S1c) for a given bin of S1c, and the summation is over the total number of S1c bins. The width of the S1c bins was chosen to be 1 phd. Both terms are different representations of the total deviation of the band means; the first term vanishes for a perfect match, assuming that any noise is Gaussian, and the second term is the quadrature sum of the deviations, and was added in the event that the first term fell into a false minimum value. We note here that $\Delta_\mu$ ignores the deviation in the Gaussian widths of the ER bands, as changes in the mean yields model had minimal effects on the band widths. 

\begin{figure}[h!]
    \centering
    \includegraphics[width=0.45\textwidth]{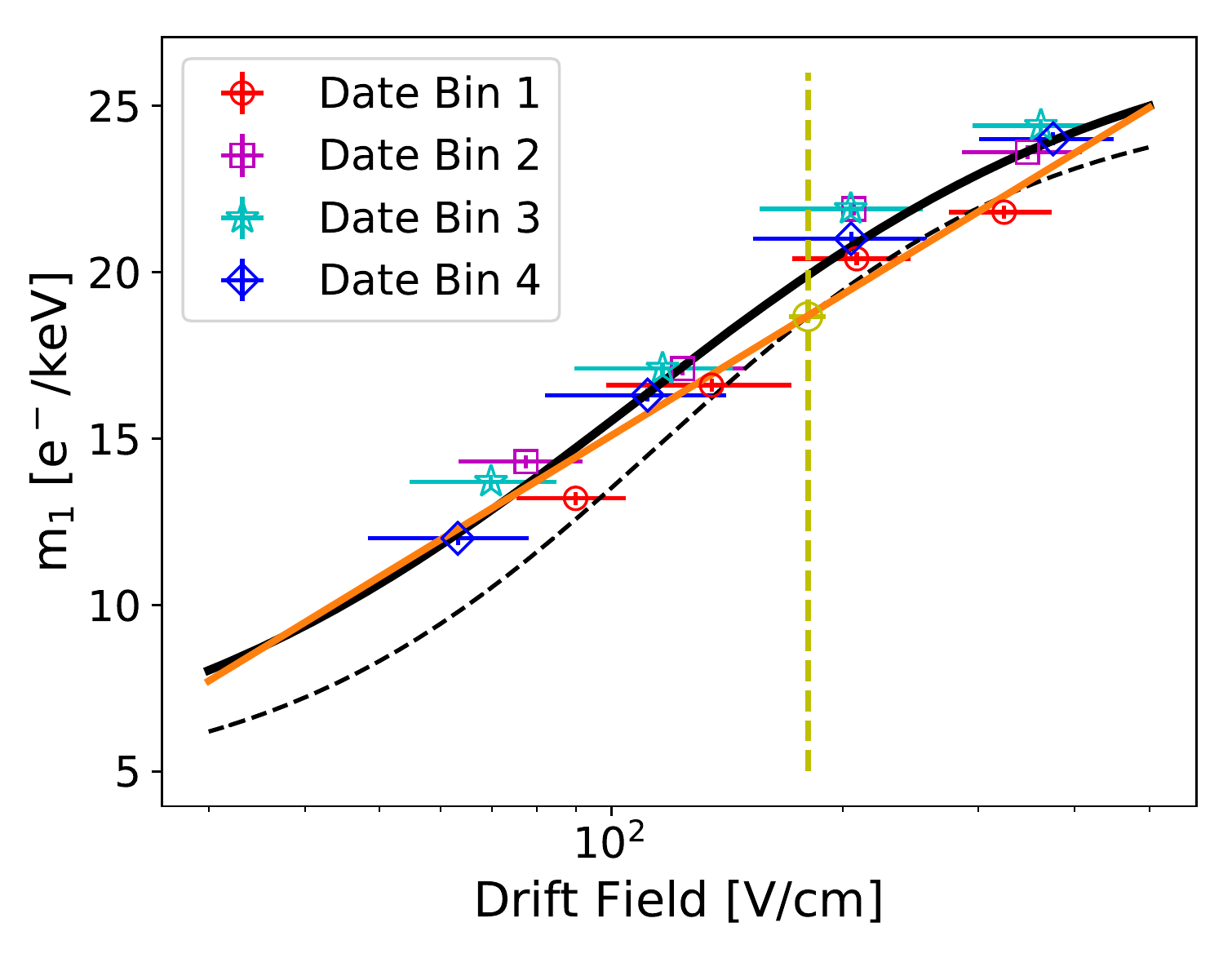}
    \caption{Best fit $m_1$ values for each bin of drift time associated with the four date bins of WS2014-16, along with the WS2013 best-fit value circled in yellow at 180 V/cm (also highlighted by the vertical yellow dashed line). The dashed black line represents the original $m_1$ model from Ref~\cite{JonThesis}. The solid black line shows the average sigmoidal best fit to the WS2014-16 data, which disagreed with the result of the WS2013 data. Because WS2014-16 data experienced complex detector effects and the WS2103 data had very little field variation, the WS2013 $m_1$ value was used to constrain the fit. A logarithmic function was chosen to split the differences given the constraint, and this final function is shown in orange.}
    \label{m1}
\end{figure}

Figure~\ref{m1} shows the resultant best fit $m_1$ found by minimizing $\Delta_\mu$ using ${}^3$H data from each of the 16 date-drift bins of WS2014-16, as well as for WS2013. The original $m_1$ model used a field-dependent sigmoid and agreed with the WS2013 data, and the WS2013 data did not experience significant field variation. Therefore, the WS2013 best fit value was used to constrain the final field-dependence of $m_1$. However, re-tuning the $m_1$ model after including the WS2014-16 data did not comply with the WS2013 constraint. Therefore, the functional form was changed to be linear in log-field space to force agreement with WS2013 while simultaneously splitting the differences seen in the WS2014-16 data. With the simplifications made to model the LUX detector during WS2014-16 inside the NEST v2.0 framework, this additional emphasis on the WS2013 data was necessary, since it was more straightforward to fully model the detector conditions from WS2013.

\begin{figure}[h!]
  \centering
  \includegraphics[width=0.48\textwidth]{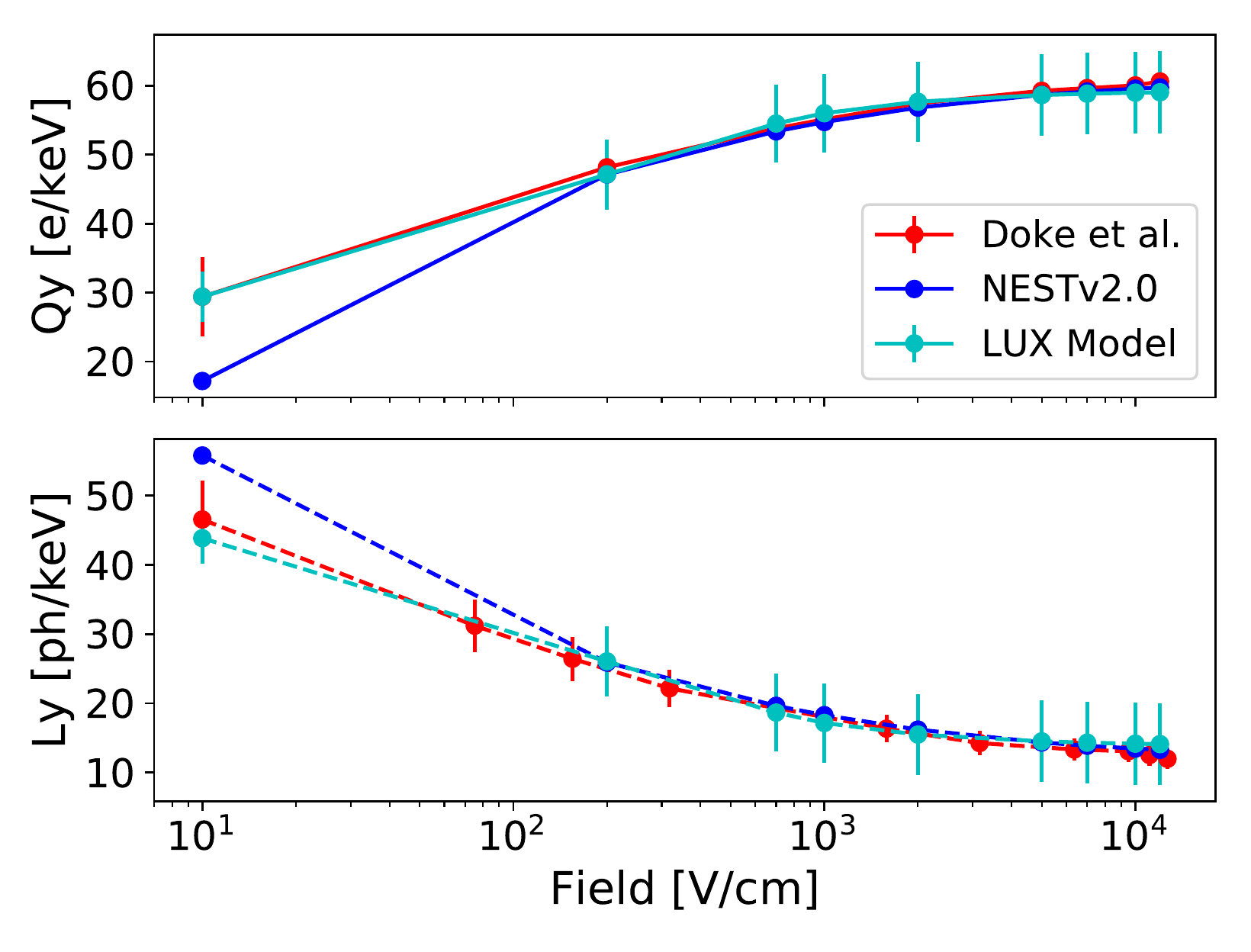}%
  \caption{(Top) Comparison to the Doke \textit{et al.} \cite{Doke} charge yield data from 976 keV Compton scatters with the public version of NEST v2.0 and the optimized LUX ER Model. This data set was used to set a limit on the new $m_1$ model, at fields much larger than seen in any of the LUX acquisitions. The change to the $m_1$ and $m_5$ models also created a much better agreement with the low field data from this data set. (Bottom) Comparison to the Doke \textit{et al.} light yield data with the public version of NEST v2.0 and the optimized LUX ER Model. This comparison highlights the anti-correlated behavior of light and charge yields, as NEST calculates light yields by subtracting charge yields from the total quanta produced in an energy deposition. }
\label{doke}
\end{figure}

This new functional form for $m_1$ grows infinitely with field strength, however, data from Doke \textit{et al.}~\cite{Doke} of measured charge yields from 976 keV Compton scatters from $^{207}$Bi suggest that $m_1$ reaches a maximum at large fields. Fitting to this data, a high-field asymptote was added to the $m_1$ model, and the comparisons to the $^{207}$Bi light and charge yields can be seen in Figure~\ref{doke}. Although Compton scatters are $\gamma$ ER, the difference in yields between $\gamma$ and $\beta$ interactions in LXe is likely due to the energy-dependent photoabsorption component of $\gamma$ interactions. Data from Compton scatters should not have a significant photoabsorption component, thus, Compton scatters should largely mimic $\beta$ interactions \cite{jonC14,nest2013}. While the field magnitudes where this cut-off is necessary are irrelevant to LUX data, this was included to allow extrapolation beyond LUX's conditions for comparison with other detectors. 

Note that as the amount of deposited energy becomes large, Equation 2.1 reduces to $Qy \approx m_1 + m_5$. Therefore, changing $m_1$ effectively alters the role of $m_5$ in the high-energy regime. For this reason, $m_5$ was constrained using the theoretical maximum charge yield. As $(n_e + n_\gamma) = E/W$, where $n_e$ and  $n_\gamma$ are the total numbers of produced electrons and photons that leave the interaction site. Then, $Qy \equiv n_e/E = \frac{1}{W}\cdot[1 + \frac{n_\gamma}{n_e}]^{-1}$. The maximum charge yield occurs when the probability of recombination vanishes, causing $n_\gamma/n_e = \alpha$, where $\alpha$ is the ratio between excited and ionized xenon atoms and is constant  for electronic recoils at the relevant energy scale for this analysis~\cite{nest1}. Therefore at large enough energies, $m_1$ + $m_5$ $\approx$ $\frac{1}{W}\cdot[1 + \alpha]^{-1}$. After changing the high energy asymptote of $Qy$, the shape of the charge yields model in the Doke-Birks regime was altered. To account for this, the field-dependence in the original $m_{10}$ model was modified. From Ref.~\cite{JonThesis}, $m_{10}$ is expressed as a field-dependent sigmoid, and it controls the curvature of the $Qy$ model at higher energies. The $m_{10}$ sigmoid was tuned by hand to restore the behavior of the new $Qy$ model in this regime to what it was before tuning $m_5$. Because the LUX ${}^3$H data consist of low energy ER events with only moderate field strength, the changes to $m_5$ and $m_{10}$ have no effect on $m_1$ optimization, although they are necessary for constructing a well-behaved model that extrapolates to higher energy ER background events. The remaining $m_i$ were not changed. 

\section{Modeling the LUX Detector}

After tuning the $\beta$ ER yields model to better match LUX data, the focus was shifted to finding nominal $g_1$ and $g_2$ values for each date bin. While $g_1$ represents the total light collection efficiency and is a fundamental property of the TPC, $g_2$ is a product of other fundamental detector properties. Explicitly: \begin{dmath} g_2 = SE \cdot \epsilon_{ext}(\mathcal{E}_{gas}) = g_1^{gas} \cdot Y_e(\mathcal{E}_{gas}, \Delta z_{gas}) \cdot \epsilon_{ext}(\mathcal{E}_{gas}), \end{dmath} where $\epsilon_{ext}$ is the efficiency to extract liberated electrons from the interaction site; $\mathcal{E}_{gas}$ is the applied electric field in the gas layer; $SE$ is the mean number of photons produced by a single extracted electron; $g_1^{gas}$ is the collection efficiency of light produced in the gas layer, which includes the PMT quantum efficiencies; $Y_e$ is the light yield of extracted electrons in the gas layer;  and $\Delta z_{gas}$ is the height of the gas layer. Both $SE$ and $g_2$ were measured multiple times throughout the course of LUX's science runs. With these, it was possible to provide NEST with starting values for $g_1^{gas}$ and $\mathcal{E}_{gas}$ after calculating the extraction efficiency using methods reported in \cite{pixey1}. For finding the best-fit values of $g_2$, $g_1^{gas}$ was chosen to be the free parameter, and the $\mathcal{E}_{gas}$ values for each date bin were left untouched. These best-fit gain factors could then be used to create the continuous temporal $g_1$ and $g_2$ for use in LLAMA's temporal interpolation.

Also, with the simplifications made to the WS2014-16 electric field maps for use in the NEST v2.0 framework, it was expected that the mean fields used in each horizontal slice of drift time would need to slightly shift from the mean values of the field maps. As mentioned in the preceding section, each drift bin was provided a range of electric fields that are associated with the mean and standard deviation of the field magnitudes in that drift bin from the relevant field map. For each modeled interaction, NEST selects the electric field from that range using a uniform random distribution.  To find effective fields that allowed for proper reconstruction of the S1c-log(S2c/S1c) bands in this simplified framework, the mean field of each drift bin was treated as a free parameter. However, the mean field value in each bin was not allowed to deviate from the original value by more than one standard deviation.  The widths of the field ranges were left fixed. This method was also used for the four drift bins associated with the ${}^{14}$C acquisition. These effective fields can provide a weighting of the field maps in LLAMA to assist in data reproduction by fitting the best-fit field means with a continuous function of drift time. The WS2013 data did not require this spatial binning since significant field fringing was not observed during the first science run. Thus, a WS2013 effective field was not required and the full 3D position dependence was implemented. 

Focusing on the WS2014-16 ${}^3$H data and the ${}^{14}$C acquisition, the deviation between the band means could be minimized for each bin of drift time. This was done by calculating $\Delta_\mu$ from Eqn.~2.2 while simultaneously varying $g_1$, $g_1^{gas}$, and the mean electric field, $\mathcal{E}$, for each drift bin. However, it was necessary to force $g_1$ and $g_1^{gas}$ to take only one set of values for each date bin and to not differ across the fiducial volumes. This is because the position reconstruction of the S1 and S2 pulses should already take the position dependence of $g_1$ and $g_2$ into account. Therefore, the average deviation for all drift bins in a single date bin was minimized. Explicitly for a single date bin,  \begin{dmath}  \Delta_\mu^{'} = \frac{1}{4} [ \Delta_{\mu 1}(g_1, g_1^{gas}, \mathcal{E}_1) + \Delta_{\mu 2}(g_1, g_1^{gas}, \mathcal{E}_2) + \Delta_{\mu 3}(g_1, g_1^{gas}, \mathcal{E}_3) + \Delta_{\mu 4}(g_1, g_1^{gas}, \mathcal{E}_4)] \end{dmath} was minimized, where the individual $\Delta_{ \mu i }$ are the aforementioned test statistics ($\Delta_\mu$) for each drift bin. This forces the resultant optimal parameters to have equal $g_1$ and $g_1^{gas}$ values for all drift time bins in a given date bin while each drift bin gets its own effective field. 

For the WS2014-16 ${}^3$H data, minimizing Eqn.~2.2 while varying the three free parameters for each date bin results in $g_1$ values of $0.0996 \pm 0.0035, 0.0994 \pm 0.0032, 0.0988 \pm 0.0023$, and $0.0976 \pm 0.0029$ for the four date bins in temporal order. Taking the best fit $g_1^{gas}$ values and converting back to $g_2$ results in $g_2$ values of $19.21 \pm 0.37, 19.66 \pm 0.26, 19.67 \pm 0.22$, and $19.96 \pm 0.28$. These results are consistent with those reported in \cite{luxRun4}. The shifts needed for the mean field in each drift bin were similar across the four date bins, thus it was possible to obtain a single set of field multipliers for the WS2014-16 data by finding the weighted average with respect to each date bin's event density. In order of increasing drift time (which is the same as decreasing electric field strength,) these averaged multipliers are $0.98, 1.16, 0.96$, and $0.86$. 

\begin{figure}[!ht]
    \centering
    \includegraphics[width=0.48\textwidth]{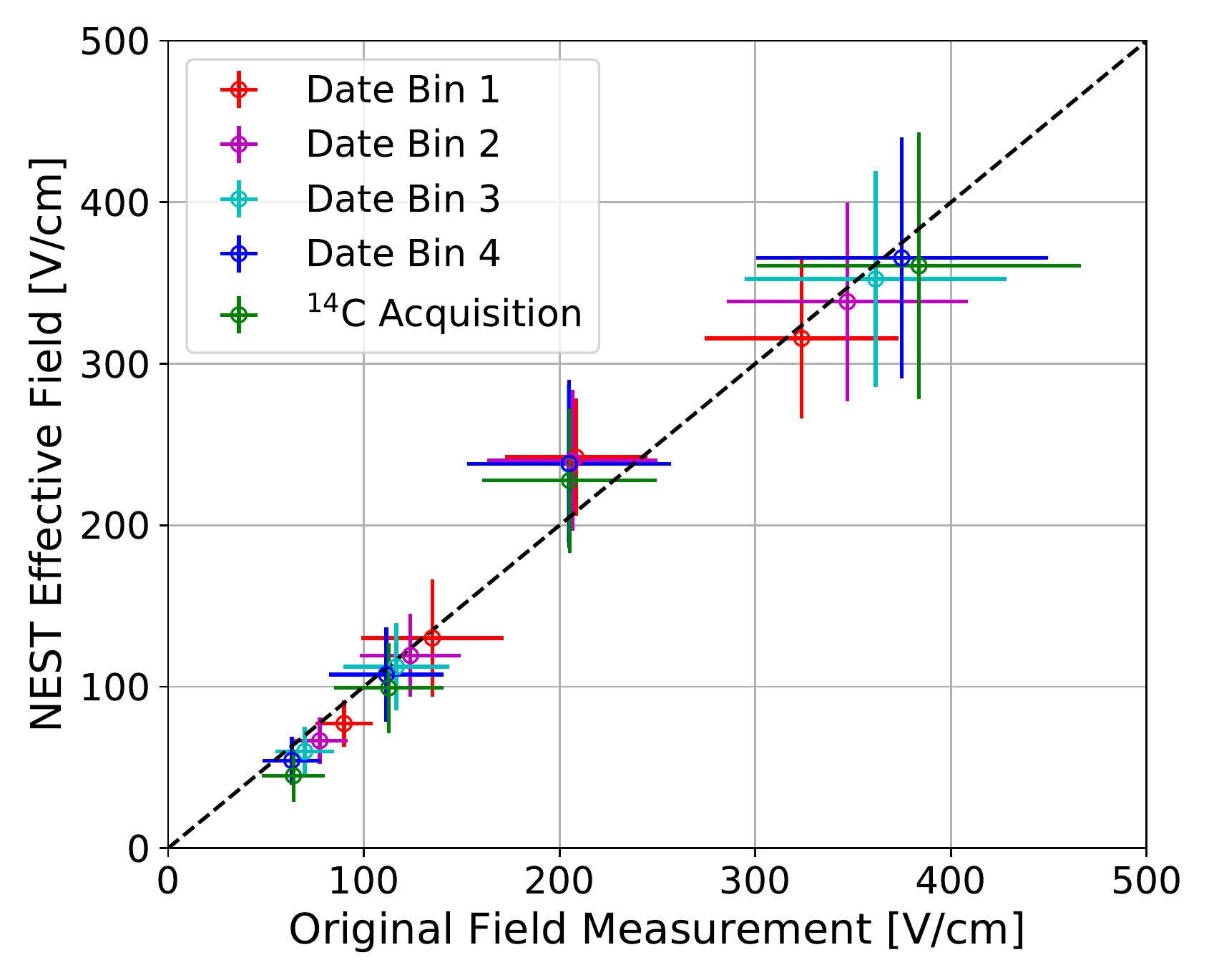}
    \caption{ Effective fields for use in the NEST v2.0 framework, compared to the original field means and widths reported in Ref.~\cite{fieldModeling}. The $y=x$ line is shown in dashed black. We note here that the mean and average deviations from the original measurements for WS2014-16 is 16\% and 9\%, which are more than a factor of two better than previously reported using older versions of NEST. Including the deviations of the ${}^{14}$C effective fields, the maximum and average deviations are 30\% and 10\%, which are still a large improvement compared to those previously reported that did not include ${}^{14}$C data. An effective field value for the WS2013 data was not required since WS2013 did not experience significant variation in the electric field.}
    \label{fieldPlot}
\end{figure}

For the ${}^{14}$C date bin, the best fit $g_1$ and calculated best fit $g_2$ are $0.0931 \pm 0.0005$ and $18.12 \pm 0.09$, respectively, which are in agreement with those reported in \cite{jonC14}. The uncertainties on each $g_1$ and $g_1^{gas}$ value are the standard deviations of the best fit $g_1$ and $g_1^{gas}$ values for the four drift bins in the given date bin. The $g_1^{gas}$ uncertainties were converted into uncertainties on $g_2$ using error propagation with Eqn.~2.1. The effective field multipliers that minimize the deviation between the ${}^{14}$C comparisons are 0.94, 1.11, 0.88, and 0.70 in order of increasing drift time. Figure~\ref{fieldPlot} shows the values of the effective fields compared to the original field means for each of the twenty drift bins.

For reproducing the WS2013 ${}^3$H data, which did not require an effective electric field, the optimal $g_1$ and $g_2$ values are 0.1165 and 13.18. While the value for $g_1$ is in agreement with that previously reported in \cite{tritium}, this $g_2$ value is increased by 1.4$\sigma$. This suggests a possible over-estimation of the previously reported ${}^3$H charge yields. All of the remaining detector parameters used in the NEST framework for WS2013, WS2014-16, and the Post-WS ${}^{14}$C acquisition are included in Table~\ref{detectorParams}.

\begin{table*}
\scriptsize
\centering
\caption{Remaining parameters used in for modeling the LUX detector in the NEST v2.0 framework for each time period. In the entries where four values are provided in WS2014-16, each value corresponds to one of the four date bins.}
\smallskip
\begin{tabular}{|c|c|c|c|}
\hline
 &  WS2013 & WS2014-16 & Post-WS \\
\hline
Primary Scintillation (S1) Parameters & & & \\
\hline
Single Photoelectron Resolution & 0.37 & 0.37 & 0.37 \\
Single Photoelectron Threshold [phe] & 0.38 & 0.38 & 0.38 \\
Single Photoelectron Efficiency & 1 & 1 & 1 \\
Gaussian Baseline Noise [phe] & -0.01 $\pm$ 0.08 & -0.01 $\pm$ 0.08 & -0.01 $\pm$ 0.08 \\
Double Photoelectron Emission Probability & 0.173 & 0.173 & 0.173 \\
\hline
Ionization and Secondary Scintillation (S2) Parameters & & & \\
\hline
g$_1^{gas}$ & 0.1 & 0.0852, 0.0928, 0.0898, 0.0854 & 0.0769 \\
Single Electron Size Fano-like Factor & 3.7 & 0.8, 1.95, 2.49, 2.67 & 2.6 \\
S2 Threshold [phe] & 192.3 & 234.6 & 234.6 \\
Extraction Region Field [kV/cm] & 6.55 & 8.134, 7.95, 8.047, 8.087 & 8.269 \\
Electron Lifetime [$\mu$s] & 800 & 735, 947, 871, 871 & 947 \\
\hline
Thermodynamic Properties & & & \\
\hline
Temperature [K] & 173 & 177 & 177 \\
Gas Pressure [bar] & 1.57 & 1.95 & 1.95 \\
\hline
Geometric Parameters & & & \\
\hline
Minimum Drift Time [$\mu$s] & 38 & 40 & 40 \\
Maximum Drift Time [$\mu$s] & 305 & 300 & 300 \\
Fiducial Radius [mm] & 180 & 200 & 200 \\
Detector Radius [mm] & 235 & 235 & 235 \\
LXe-GXe Border [mm] & 544.95 & 544.95 & 544.95 \\
Anode Level [mm] & 549.2 & 549.2 & 549.2 \\
Gate Level [mm] & 539.2 & 539.2 & 539.2 \\
Cathode Level [mm] & 56 & 56 & 55.9 \\
\hline
\end{tabular}
\label{detectorParams}
\end{table*}

\section{Modeling Yield Fluctuations}

All of the aforementioned model optimizations compared only the deviation between the band means of the LUX data and simulated events in S1c-log(S2c/S1c) space, ignoring comparisons between the Gaussian band widths. This was because changes in the mean yields model, $g_1$, $g_2$, and the mean drift field resulted in little change in the band widths. However, after the improved comparison between the band means, disagreements remained between the band widths, especially for data with larger S1s. This suggested that NEST's fluctuations models needed their own optimization to reproduce LUX data, and careful consideration of the possible contributions to fluctuations in S1 and S2 pulses was taken. Three sources of pulse fluctuations were identified: statistical quantum fluctuations in the total produced quanta, $n_q$, recombination fluctuations from ionized electrons recombining with xenon atoms before extraction, and additional random noise in pulse areas due to experimental detector effects. The first of these creates a correlated fluctuation in the S1 and S2 areas; as the total yield fluctuates, the number of electrons and photons should fluctuate in the same direction relative to the mean value. The second source, recombination fluctuations, are anti-correlated, as a recombining electron will excite a xenon atom, eventually leading to de-excitation and possible photon production. These two sources are fundamental to the atomic physics of energy depositions in xenon targets. The last source of fluctuations, noise from detector effects, can be different for any experiment and are not currently included in NEST v2.0. These fluctuations are expected to have uncorrelated impacts on S1 and S2 size.

Comparing the calculated energy resolution in NEST v2.0  with those measured in WS2013 and reported by LUX in Ref.~\cite{SignalYields}, the discrepancy is not significant compared to the differences seen in the simulated and measured S1c-log(S2c/S1c) band widths. Because recombination fluctuations have no significant effect on the combined energy resolution, this suggests that the correlated statistical fluctuations model in NEST v2.0 did not require any additional tuning for the LUX yields model.  Therefore, we set our focus on tuning the NEST ER recombination model to reduce the remaining deviation in the S1c and S2c distributions. 

Historically, modeling recombination as a binomial process - either an electron recombines or it does not - has been unable to fully explain the observed anti-correlated yield fluctuations in light and charge yields \cite{Conti}. Reported by LUX in Ref.~\cite{SignalYields}, recombination fluctuations on the number of produced ions, $N_{ions}$, have a variance of the form: \begin{equation} \sigma^2 = r(1-r)N_{ions} + \omega^2N_{ions}^2. \end{equation} The first term can be recognized as a binomial term with recombination probability, $r$, while the second term is the non-binomial correction weighted by a field-dependent parameter, $\omega$. This is also the same form that recombination fluctuations take in the NEST v2.0 framework, with the weighting parameter, $\omega$, represented by a field-dependent quadratic function of recombination probability which makes use of three free parameters. For optimizing the LUX yields model, NEST's $\omega$ representation was adopted, but simplifications were made by reducing the number of free parameters to  a single field-dependent variable. Specifically, \begin{dmath}
\omega = -4c(r - \frac{1}{2})^2 + c,
\end{dmath} where $c$ is the free parameter. This is useful as it forces the non-binomial variance contributions to vanish when $r$ becomes zero or unity, and reaches a maximum when the probabilities for recombination and escape are equal. In addition, while NEST v2.0 models recombination fluctuations as a Gaussian process, data reveals that the final S2 distribution is more accurately represented by a skewed Gaussian distribution~\cite{zepIII}. Therefore, the LUX yields model expanded upon the NEST v2.0 recombination model to incorporate skew in the final S2 distributions.

To begin the optimization process, a new test statistic was required, as the choice for the band mean optimization (Eqn.~2.2) was agnostic to the changes in the band widths. Therefore,
\begin{dmath} \Delta = \frac{1}{N} [\sum|\mu_{_{NEST}} - \mu_{_{data}}| + \sum|\sigma_{_{NEST}} - \sigma_{_{data}}|] \end{dmath} 
was chosen as the optimization test statistic: the summed average deviations of the band means and band widths. Here, $\mu$ is the mean value of log(S2c/S1c) for a given S1c bin, while $\sigma$ is the standard deviation; $\mu$ and $\sigma$ were found using Gaussian fits. As before, the summation takes place over the number of S1c bins. Although the band means have already been optimized, including the deviation between the means in $\Delta$ prevents any changes in the fluctuations adversely impacting the previously accomplished improvements.

\begin{figure}[h]
    \centering
    \includegraphics[width=0.45\textwidth]{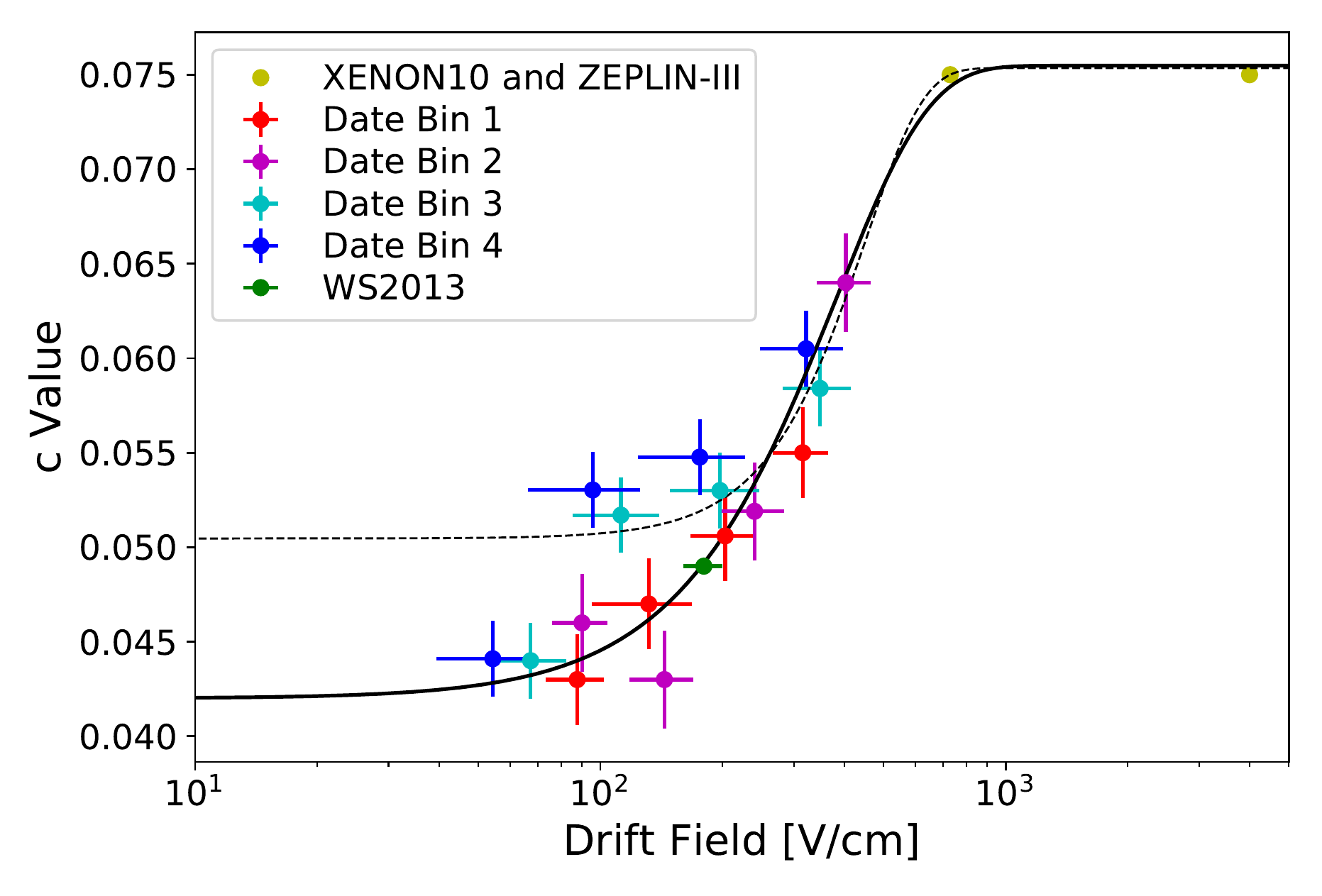}
    \caption{Best-fit $c$ values as a function of field shown as a solid black line, found by fitting to the WS2013 and WS2014-16 ${}^3$H data, as well as XENON10 and ZEPLIN-III data. Because of the similar values for XENON10 and ZEPLIN-III that suggest the need for a high-field asymptote, a sigmoid was chosen as the best fit function. This is the same functional form as used originally for NEST v2.0, indicated by the dashed black line.
    The fit to the WS2014-16 data was constrained using the WS2013 result, as detector effects were not significant during this data collection campaign. }
    \label{C}
\end{figure}

Similar to the approach taken in the mean yields optimization, the free parameters were found that minimized the test statistic for each of the WS2013 and WS2014-2016 drift bins using ${}^3$H data.  Figure~\ref{C} shows the best fit values of $c$ for the 17 LUX bins, and also includes data points from XENON10 and ZEPLIN-III at higher fields than LUX \cite{XENON10, zepIII}. The inclusion of data from these different experiments allows for extrapolation of the recombination model beyond LUX's data-taking conditions. These data also indicate asymptotic behavior in $c$ as the drift field increases. Therefore, $c$ was modeled as a field-dependent sigmoid and was fit to the WS2014-16 data. As with the band mean optimization, the WS2013 result was used to constrain the fit because the first science run did not experience significant temporal and spatial detector effects. 

\begin{figure}[!ht]
    \centering
    \includegraphics[width=0.55\textwidth]{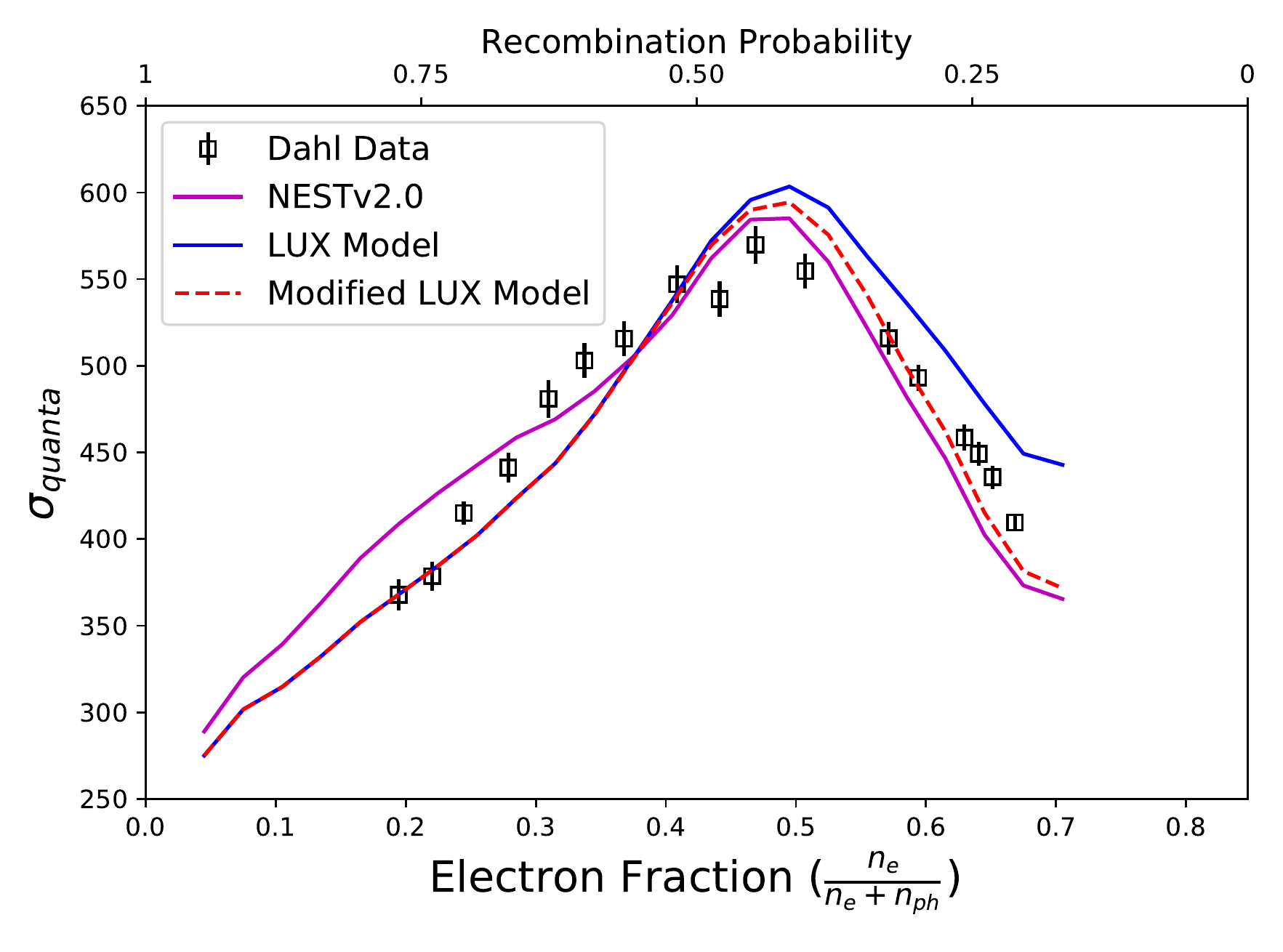}
    \caption{Recombination fluctuations as a function of the ratio of electrons to total quanta compared with the NEST v2.0 model (magenta) and the LUX yields model before and after modification with an error function (solid blue and dashed red, respectively). The electron fraction is equal to $(1-r)/(\alpha + 1)$, where $r$ is the recombination probability and $\alpha$ is the ratio of excitons to ions produced in the original energy deposition.}
    \label{dahlCompare}
\end{figure}

The data used in the $\omega$ model thus far was limited in range of the recombination probability, and as $\omega$ depends directly on $r$, it was important for extrapolation purposes to check the behavior of the model for a large spread of recombination probabilities. Data provided in a dissertation by C.E. Dahl  \cite{DahlThesis} provides recombination fluctuation data for a wide range of electron fraction (defined as $n_e/(n_e + n_{ph})$), which is proportional to $(1-r)$. Because the model should provide the true recombination fluctuations for an event, it is favorable that the fluctuations in the model are underestimated instead of being overestimated compared to data, as data may contain unknown sources of noise. With this in mind, initial comparisons show that the new LUX yields model performs better when recombination is more probable than escape, but the original NEST v2.0 model performs better at lower values of the recombination probability. Because of this, the LUX yields $\omega$ model was weighted with a recombination-dependent error function, helping to improve the model for lower recombination probabilities. Figure~\ref{dahlCompare} shows this data compared to the NEST v2.0 model and the LUX model, both before and after this final modification. The effects of this change on the LUX data comparisons were negligible for all of the relevant fields and energies, so re-optimization of the LUX model parameters was not necessary.

\section{Modeling Detector Noise in S1 and S2 Pulses}

With the newly finished recombination model optimized for a wide range of recombination probabilities, the last source of yield fluctuations to address were those from detector effects. It is understandable that this would be necessary for the WS2014-16 and ${}^{14}$C data, as the field non-uniformity and time-dependent light collection efficiencies made proper simulation of the LUX data difficult. These fluctuations were modeled as corrections on the S1 or S2 pulse area, $A$, using a Gaussian smearing with a standard deviation, $\lambda A$, where $\lambda$ is a free parameter that encapsulates complicated or unknown detector effects that impact the area of the detected S1 or S2, similar to the techniques used in Ref.~\cite{DahlThesis}. For simplicity, $\lambda$ was chosen to be identical for smearing of both S1 and S2 pulses.

Best-fit $\lambda$ values were found for the ${}^3$H data from WS2013 and each WS2014-16 date bin, as well as for the ${}^{14}$C acquisition. Because the effects of this method of modeling detector noise increase as the pulse areas increase, the test statistic (Eqn.~4.3) was minimized using data out to the largest S1s available for a given date bin: 115 phd for WS2013 ${}^3$H data and between 80 and 100 phd for WS2014-16 ${}^3$H data. The WS2014-16 S1 range is less than the maximum S1 for WS2013 because $g_1$ decreased and was changing during the second science run. For ${}^{14}$C, only data with S1 pulses smaller than 300 phd were used, as the number of events becomes too few beyond this limit. For WS2013, the four WS2014-16 date bins, and the ${}^{14}$C acquisition, the best fit values of $\lambda$ are: $1.4 \pm 0.7\%, 1.4 \pm 0.7\%,  2.3 \pm 0.4\%,  3.0 \pm 0.6\%,  4.4 \pm 0.6\%$,  and $7.0 \pm 0.1\%$, respectively. This is consistent with the expectation that detector effects were steadily increasing over time. As with the gain factors, $g_1$ and $g_2$, a smooth temporal fit to these $\lambda$ values can be added into LLAMA for accurate reproduction of the band widths for WS2014-16 and the ${}^{14}$C acquisition.

\section{Results}

\begin{figure}
    \centering
    \includegraphics[width=0.5\textwidth]{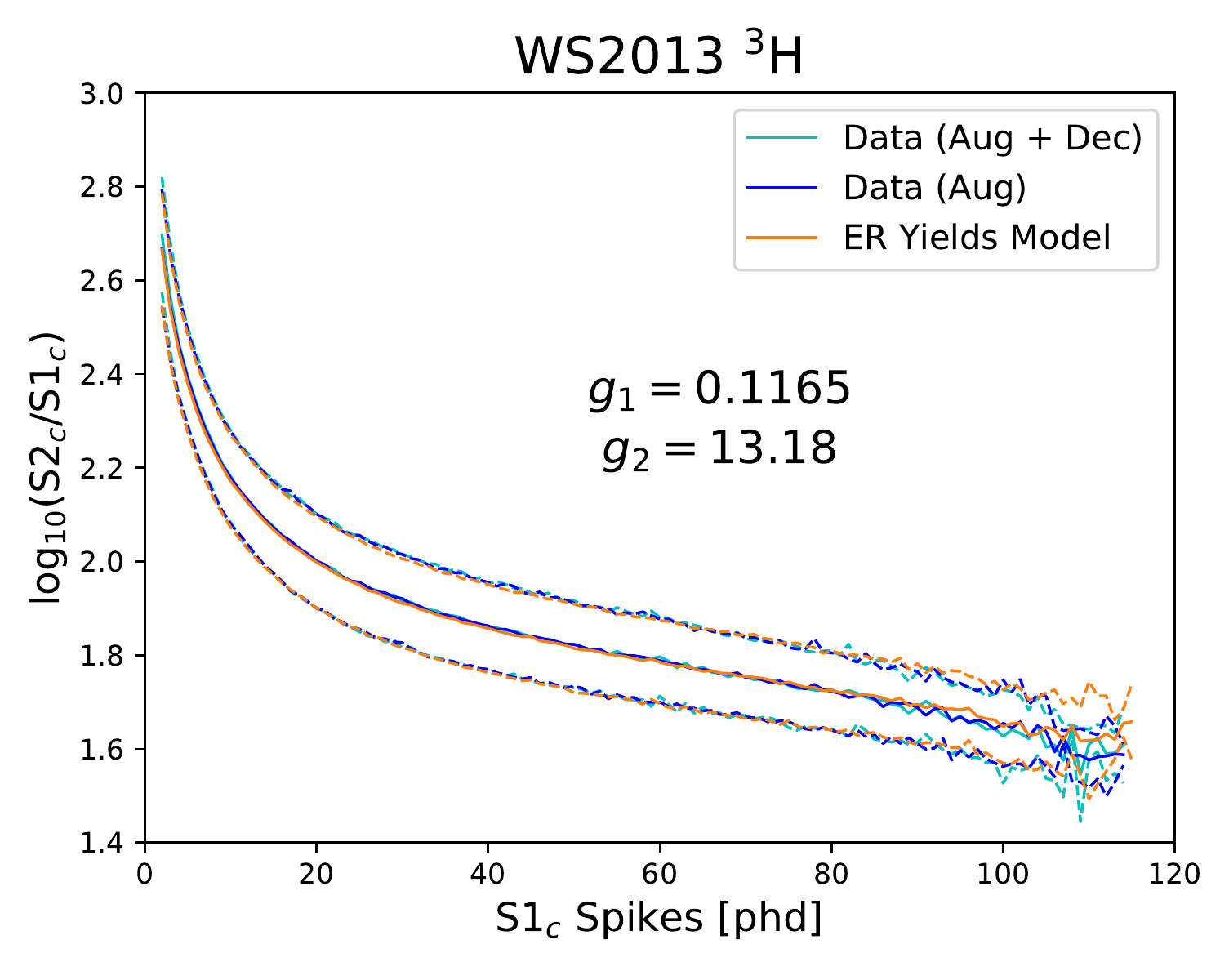}
    \caption{Comparison of the final LUX ER yields model using the NEST v2.0 framework with two different sets of LUX WS2013 ${}^3$H data. The solid lines represent the Gaussian means of log(S2c/S1c) values inside each S1 bin. Dashed lines show the Gaussian widths of each S1 bin. }
    \label{run3}
\end{figure}

\begin{figure*}
    \centering
    \includegraphics[width=\textwidth]{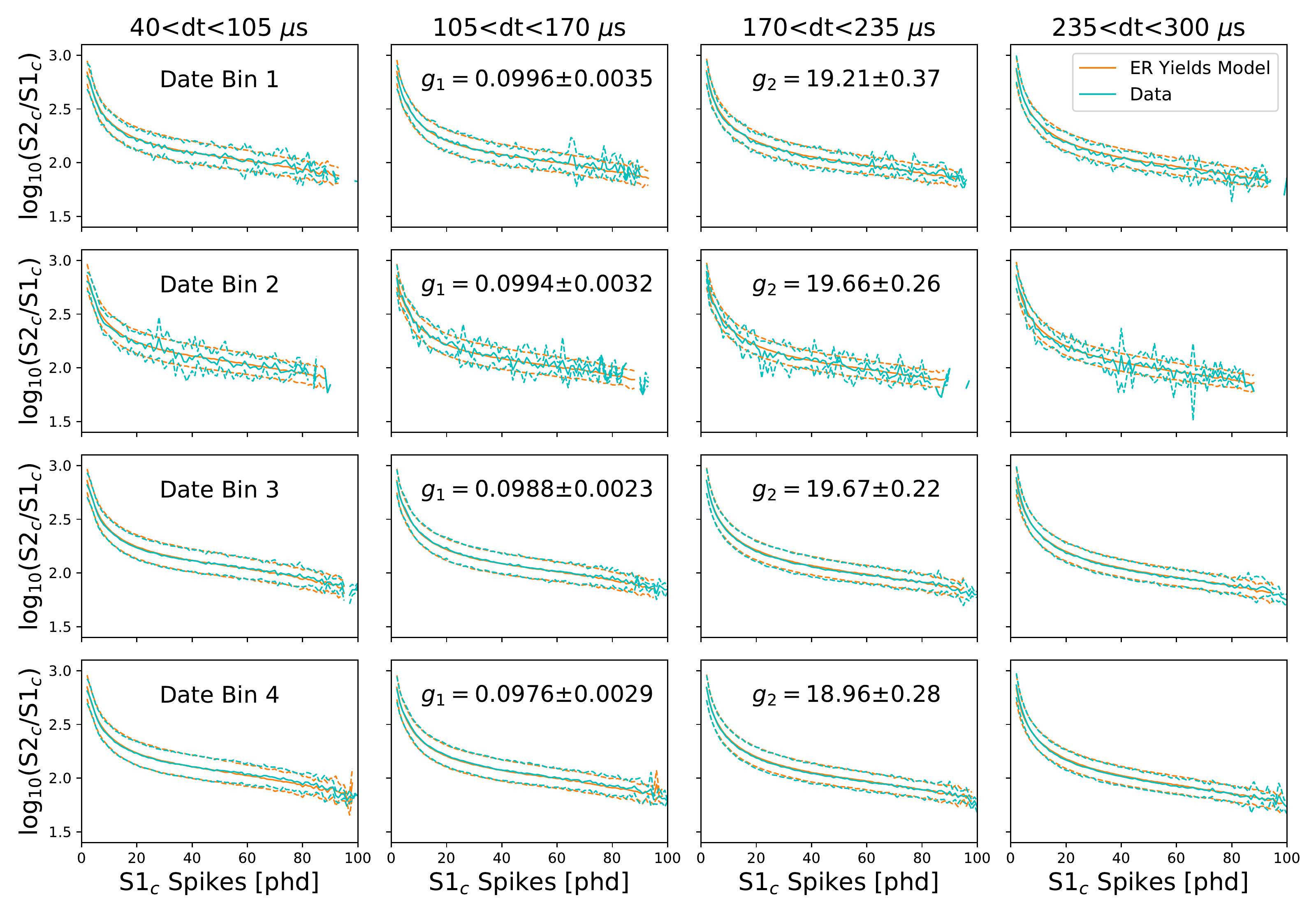}
    \caption{Comparison of the final LUX ER yields model using the NEST v2.0 framework with ${}^3$H data from each of the WS2014-16 drift bins. Each row shows one of the four date bins with the best-fit g$_1$ and g$_2$ values, and each column represents one of the four bins of drift time. The solid lines represent the Gaussian means of log(S2c/S1c) values inside each S1 bin. Dashed lines show the Gaussian widths of each S1 bin.}
    \label{run4}
\end{figure*}

\begin{figure}
    \centering
    \includegraphics[width=0.58\textwidth]{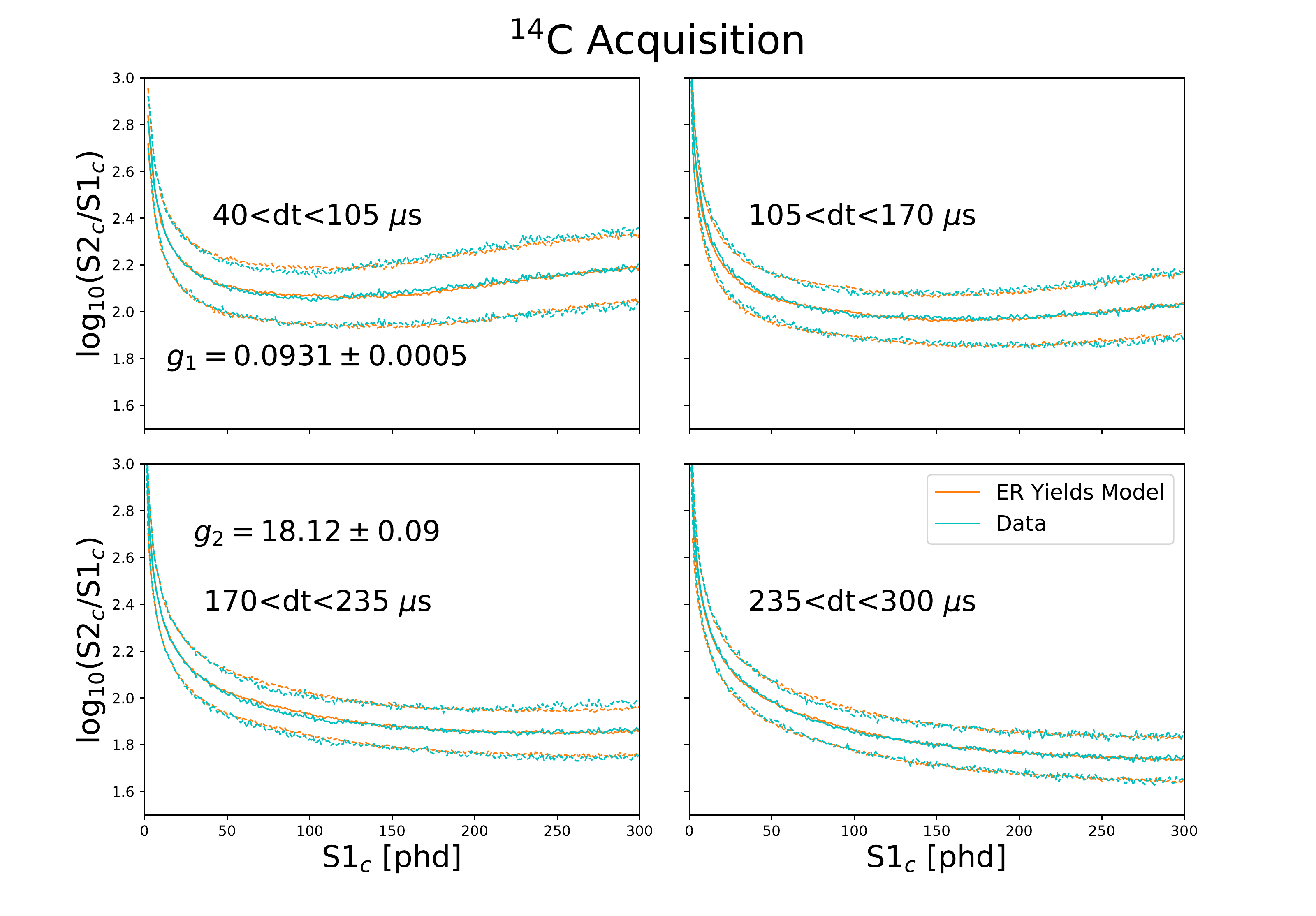}
    \caption{Comparison of NESTv2 and LUX post-Run04 ${}^{14}$C data. Similarly to the Run04 data, ${}^{14}$C data is split into four spatial bins of equal drift time. The solid lines represent the Gaussian means of log(S2c/S1c) values inside each S1 bin. Dashed lines show the Gaussian widths of each S1 bin. Note that units of S1s are expressed in either S1 spikes or pulse areas in phd. Typically, S1s less than approximately 120 phd will be expressed in terms of spikes, and anything larger will be expressed in terms of standard S1 pulse areas.}
    \label{c14}
\end{figure}

\begin{table*}[ht]
\centering
\caption{Average signed deviation between the NEST simulation and LUX data for each acquisition period for both the means and widths, calculated as $(X_{NEST} - X_{LUX})/X_{NEST}$, since the LUX data was limited in statistics and dividing by a data point with a significant downward fluctuation would inflate the calculated deviations. The sub-percent deviations along with the variation between positive and negative values indicate that there is no systematic offset between the simulation and the data. The large deviation in the widths for the lowest field bin in Date Bin 2 is due to low statistics in the data, and this can be seen in the corresponding pane of Figure~\ref{run4}.}
\footnotesize
\begin{tabular}{|c|c|c|c|c|}
\hline
LUX Acquisition & Drift Time Bin & Max S1 & Avg Deviation: Means & Avg Deviation: Widths \\
\hline
\hline
Run 3 & - & 105 & 0.01\% & -0.31\% \\
\hline
\multirow{4}{*}{ Run 4: Date Bin 1 } & 1 & 70 & 0.11\% & 2.06\% \\
& 2 & 70 & -0.21\% & 2.19\% \\
& 3 & 70 & 0.37\% & -1.12\% \\
& 4 & 70 & 0.65\% & 2.13\% \\
\hline
\multirow{4}{*}{ Run 4: Date Bin 2 } & 1 & 70 & 0.33\% & -5.77\% \\
& 2 & 70 & -0.25\% & -2.94\% \\
& 3 & 70 & 0.39\% & -2.85\% \\
& 4 & 70 & 0.34\% & 16.10\% \\
\hline
\multirow{4}{*}{ Run 4: Date Bin 3 } & 1 & 70 & 0.11\% & 1.83\% \\
& 2 & 70 & -0.22\% & -0.40\% \\
& 3 & 70 & 0.39\% & -4.51\% \\
& 4 & 70 & 0.20\% & -0.90\% \\
\hline
\multirow{4}{*}{ Run 4: Date Bin 4 } & 1 & 70 & -0.09\% & 1.62\% \\
& 2 & 70 & -0.30\% & 0.12\% \\
& 3 & 70 & 0.38\% & -4.71\% \\
& 4 & 70 & 0.35\% & 0.91\% \\
\hline
\multirow{4}{*}{ Run 4: C-14 Acquisition } & 1 & 300 & -0.10\% & -0.31\% \\
& 2 & 300 & -0.33\% & -1.25\% \\
& 3 & 300 & 0.01\% & -5.47\% \\
& 4 & 300 & -0.22\% & 3.72\% \\
\hline
\hline
Averages & - & - & 0.09\% & -1.79\% \\
\hline
\end{tabular}
\label{devTable}
\end{table*}

After tuning the NEST v2.0 mean ER charge yields and fluctuations models and finding optimal light collection efficiencies and effective drift fields, final comparisons can be made with the LUX data. Figures~\ref{run3},~\ref{run4}, and~\ref{c14} show the comparisons of the final LUX yields model to the WS2013 ${}^3$H data, each of the 16 drift bins associated with the WS2014-16 ${}^3$H data, and the four ${}^{14}$C drift bins.  The simulated bands were created with increased statistics to reduce the effects of Poisson fluctuations obscuring the final quantitative and qualitative results.

\begin{figure}
    \centering
    \includegraphics[width=0.58\textwidth]{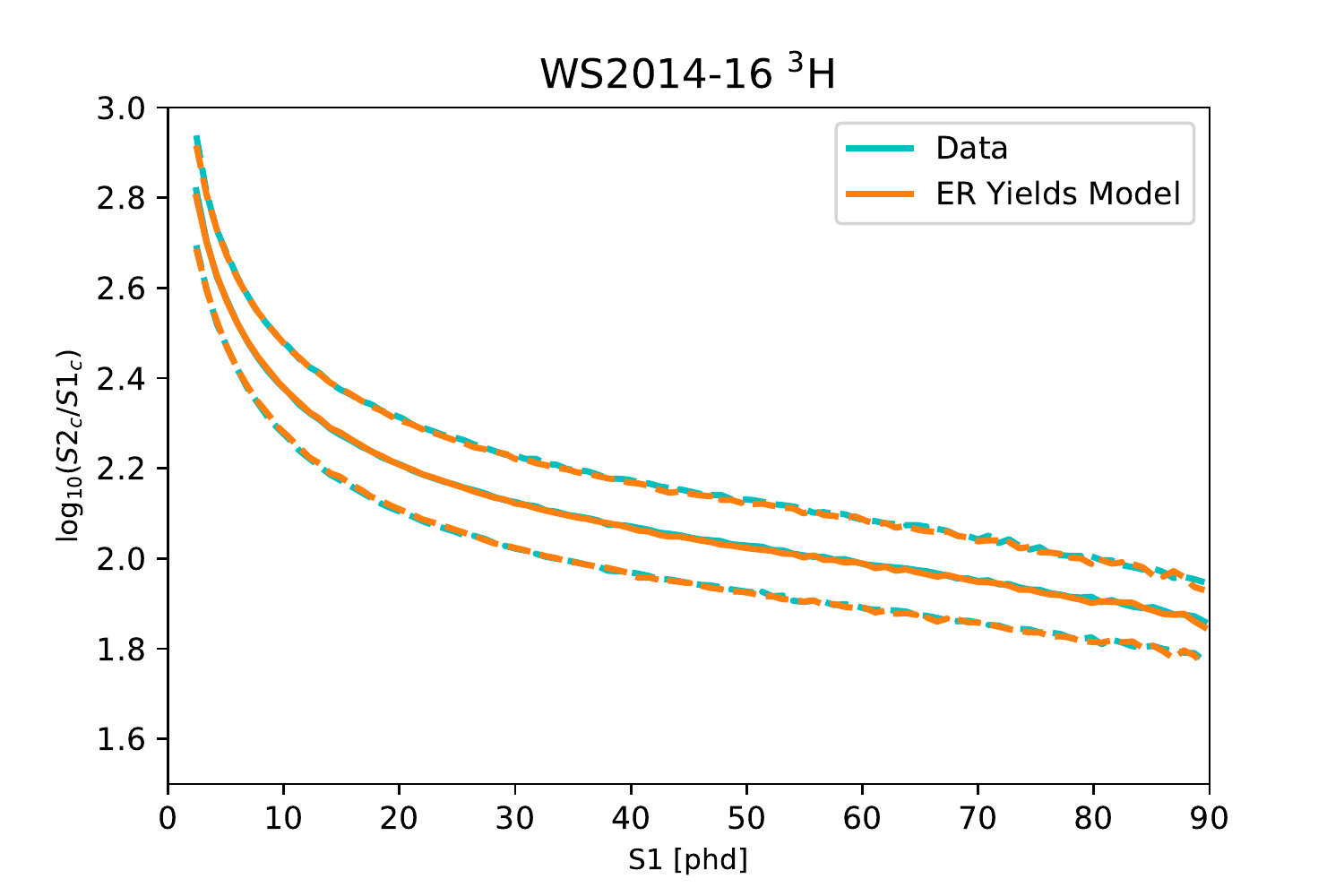}
    \caption{ Comparison of the final LUX ER yields model using LLAMA with the WS2014-16 ${}^{3}$H data. LLAMA uses the temporal dependence of g$_1$ and g$_2$ found with the four date bins. In addition, the full time-dependent and position-dependent field map is used and is weighted with the optimized effective field multipliers found with the discretized approach.}
    \label{llama4}
\end{figure}

\begin{figure}
    \centering
    \includegraphics[width=0.58\textwidth]{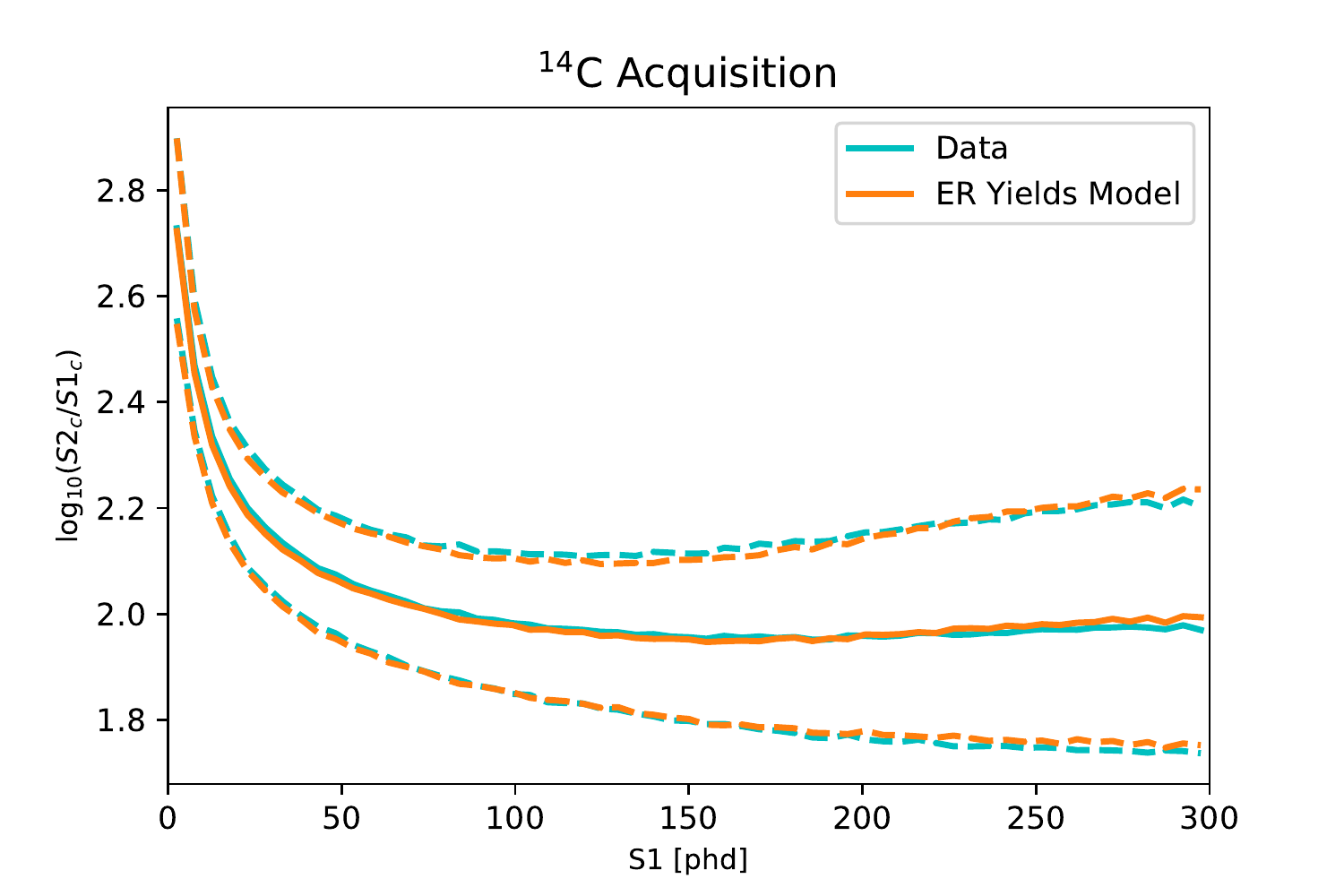}
    \caption{Comparison of the final LUX ER yields model using LLAMA with the ${}^{14}$C acquisition data. This comparison uses the full position-dependent field map and is weighted with the optimized effective field multipliers found for the ${}^{14}$C data with the discretized approach.}
    \label{llama5}
\end{figure}

In addition to this discretized approach, Figures~\ref{llama4} and~\ref{llama5} show the comparison using LLAMA for the total WS2014-16 band and the ${}^{14}$C band. Using the best-fit detector parameters and creating continuous temporal functions for each one allows LLAMA to take the newly modified NEST ER models and accurately reproduce the LUX data, despite the significant detector pathologies that affected the data. 

Table~\ref{devTable} shows the remaining net average relative deviation between each comparison. For the band means, the average deviations never become larger than 1\%, quantitatively indicating good agreement between the data and simulations. The magnitude of the net deviations for the band widths are on average larger than those for the band means. This is due to larger Poisson fluctuations and relative uncertainties in the band width calculations and measurements. 
These values also show that the remaining deviation does not appear to be systematically offset. Importantly, these final values reveal that very little additional progress can be made beyond the improvements presented here. 

\section*{Conclusion}

The final LUX backgrounds model requires an ER yields model that can mimic the data collected throughout the LUX experiment. The techniques and methods described in this paper present a detailed and robust application of NEST v2.0 for use in accurately reproducing light and charge yields from TPCs with xenon targets. With minimal tuning of the NEST v2.0 models, it was possible to reproduce LUX ER calibration data from each of the different acquisition time periods. Even with the significant field fringing and time-dependent light collection efficiencies present in the data beyond WS2013, the NEST v2.0 framework was able to accurately and efficiently encapsulate these effects with simplified geometry after binning the detector spatially and temporally, while also providing nominal values of $g_1$ and $g_2$ for each time bin and nominal effective electric field ranges. The results found using the temporal and spatial binning of the LUX detector were interpolated to create smooth functions for the relevant detector parameters. Expanding upon the NEST v2.0 framework to create LLAMA for simulating a temporally dynamic detector, this final model can accurately reproduce all LUX ER data, despite the significant detector pathologies observed in later data acquisitions. The completed LLAMA framework equipped with the final LUX ER model will be used for all future background modeling in upcoming LUX analyses.


\section*{Acknowledgments}

The research supporting this work took place in whole or in part at the Sanford Underground Research Facility (SURF) in Lead, South Dakota. Funding for this work is supported by the U.S. Department of Energy, Office of Science, Office of High Energy Physics under Contract Number DE-SC0020216. This work was also partially supported by the U.S. Department of Energy (DOE) under award numbers DE-FG02-08ER41549, DE-FG02-91ER40688, DE-FG02-95ER40917, DE-FG02-91ER40674, DE-NA0000979, DE-FG02-11ER41738, DE-SC0015535, DE-SC0006605, DE-AC02-05CH11231, DE-AC52-07NA27344, and DE-FG01-91ER40618; the U.S. National Science Foundation under award numbers PHYS-0750671, PHY-0801536, PHY-1004661, PHY-1102470, PHY-1003660, PHY-1312561, PHY-1347449; the Research Corporation grant RA0350; the Center for Ultra-low Background Experiments in the Dakotas (CUBED); and the South Dakota School of Mines and Technology (SDSMT). LIP-Coimbra acknowledges funding from Funda\c{c}\~{a}o para a Ci\^{e}ncia e a Tecnologia (FCT) through the project-grant CERN/FP/123610/2011. Imperial College and Brown University thank the UK Royal Society for travel funds under the International Exchange Scheme (IE120804). The UK groups acknowledge institutional support from Imperial College London, University College London and Edinburgh University, and from the Science \& Technology Facilities Council for PhD studentship ST/K502042/1 (AB). The University of Edinburgh is a charitable body, registered in Scotland, with registration number SC005336. 

We gratefully acknowledge the logistical and technical support and the access to laboratory infrastructure provided to us by SURF and its personnel. SURF was developed by the South Dakota Science and Technology authority, with an important philanthropic donation from T. Denny Sanford
\bibliography{list}{}

\begin{thebibliography}{10}

\bibitem{lux}
D.~S. Akerib~\textit{et al.} (LUX~Collaboration), ``{The Large Underground
  Xenon (LUX) Experiment},'' {\em Nucl. Instrum. Meth.}, vol.~A704,
  pp.~111--126, 2013.

\bibitem{luxFirst}
D.~S. Akerib~\textit{et al.} (LUX~Collaboration), ``{First Results from the LUX
  Dark Matter Experiment at the Sanford Underground Research Facility},'' {\em
  Phys. Rev. Lett.}, vol.~112, p.~091303, 2014.

\bibitem{lux2}
D.~S. Akerib~\textit{et al.} (LUX~Collaboration), ``{Improved Limits on
  Scattering of Weakly Interacting Massive Particles from Reanalysis of 2013
  LUX Data},'' {\em Phys. Rev. Lett.}, vol.~116, p.~161301, 2016.

\bibitem{luxRun4}
D.~S. Akerib~\textit{et al.} (LUX~Collaboration), ``{Results from a Search for
  Dark Matter in the Complete LUX Exposure},'' {\em Phys. Rev. Lett.},
  vol.~118, p.~021303, 2017.

\bibitem{fieldModeling}
D.~S. Akerib~\textit{et al.} (LUX~Collaboration), ``{3D Modeling of Electric
  Fields in the LUX Detector},'' {\em Journal of Instrumentation}, vol.~12,
  no.~11, 2017.

\bibitem{xenon100_1year}
E.~Aprile~\textit{et al.} (XENON~Collaboration), ``{Dark Matter Search Results
  from a One Ton-Year Exposure of XENON1T},'' {\em Phys. Rev. Lett.}, vol.~121,
  p.~111302, 2018.

\bibitem{pandaX}
X.~Cui~\textit{et al.} (PandaX-II~Collaboration), ``{Dark Matter Results from
  54-Ton-Day Exposure of PandaX-II Experiment},'' {\em Phys. Rev. Lett.},
  vol.~119, p.~181302, 2017.

\bibitem{TDR}
B.~J. Mount~\textit{et al.}, ``{LUX-ZEPLIN (LZ) Technical Design Report},''
  2017.

\bibitem{LZsens}
D.~S. Akerib~\textit{et al.} (LUX-ZEPLIN~Collaboration), ``{Projected WIMP
  Sensitivity of the LUX-ZEPLIN (LZ) Dark Matter Experiment},'' 2018.

\bibitem{axionLUX}
D.~S. Akerib~\textit{et al.} (LUX~Collaboration), ``{First Searches for Axions
  and Axionlike Particles with the LUX Experiment},'' {\em Phys. Rev. Lett.},
  vol.~118, no.~26, p.~261301, 2017.

\bibitem{mirrorDM}
D.~S. Akerib~\textit{et al.} (LUX~Collaboration), ``{First direct detection
  constraint on mirror dark matter kinetic mixing using LUX 2013 data},'' 2019.

\bibitem{tritium}
D.~S. Akerib~\textit{et al.} (LUX~Collaboration), ``{Tritium calibration of the
  LUX dark matter experiment},'' {\em Phys. Rev.}, vol.~D93, no.~7, p.~072009,
  2016.

\bibitem{jonC14}
D.~S. Akerib~\textit{et al.} (LUX~Collaboration), ``{Improved Measurements of
  the $\beta$-Decay Response of Liquid Xenon with the LUX Detector},'' 2019.

\bibitem{nestv2}
M.~Szydagis~\textit{et al.}, ``{Noble Element Simulation Technique v2.0
  (10.5281/zenodo.1314669)},'' 2018.

\bibitem{DahlThesis}
C.~E. Dahl, {\em {The physics of background discrimination in liquid xenon, and
  first results from XENON10 in the hunt for WIMP dark matter}}.
\newblock PhD thesis, Princeton U., 2009.

\bibitem{JonThesis}
J.~Balajthy, {\em {Purity Monitoring Techniques and Electronic Energy
  Deposition Properties in Liquid Xenon Time Projection Chambers}}.
\newblock PhD thesis, University of Maryland, 2018.

\bibitem{TIB}
J.~Thomas and D.~A. Imel, ``Recombination of electron-ion pairs in liquid argon
  and liquid xenon,'' {\em Phys. Rev. A}, vol.~36, pp.~614--616, 1987.

\bibitem{DB}
T.~Doke~\textit{et al.}, ``{Let Dependence of Scintillation Yields in Liquid
  Argon},'' {\em Nucl. Instrum. Meth.}, vol.~A269, pp.~291--296, 1988.

\bibitem{nest1}
M.~Szydagis~\textit{et al.}, ``{NEST}: a comprehensive model for scintillation
  yield in liquid xenon,'' {\em Journal of Instrumentation}, vol.~6, no.~10,
  p.~P10002, 2011.

\bibitem{Doke}
T.~Doke~\textit{et al.}, ``{Absolute Scintillation Yields in Liquid Argon and
  Xenon for Various Particles},'' {\em Japanese Journal of Applied Physics},
  vol.~41, no.~Part 1, No. 3A, pp.~1538--1545, 2002.

\bibitem{nest2013}
M.~Szydagis~\textit{et al.}, ``{Enhancement of {NEST} capabilities for
  simulating low-energy recoils in liquid xenon},'' {\em Journal of
  Instrumentation}, vol.~8, no.~10, p.~C10003, 2013.

\bibitem{pixey1}
B.~N.~V. Edwards~\textit{et al.}, ``Extraction efficiency of drifting electrons
  in a two-phase xenon time projection chamber,'' {\em Journal of
  Instrumentation}, vol.~13, no.~01, p.~P01005, 2018.

\bibitem{SignalYields}
D.~S. Akerib~\textit{et al.} (LUX~Collaboration), ``{Signal yields, energy
  resolution, and recombination fluctuations in liquid xenon},'' {\em Phys.
  Rev. D}, vol.~95, 2017.

\bibitem{Conti}
E.~Conti~\textit{et al.}, ``{Correlated fluctuations between luminescence and
  ionization in liquid xenon},'' {\em Phys. Rev.}, vol.~B68, p.~054201, 2003.

\bibitem{zepIII}
V.~N. Lebedenko~\textit{et al.}, ``{Result from the First Science Run of the
  ZEPLIN-III Dark Matter Search Experiment},'' {\em Phys. Rev.}, vol.~D80,
  p.~052010, 2009.

\bibitem{XENON10}
E.~Aprile {\em et~al.}, ``{Design and performance of the XENON10 dark matter
  experiment},'' {\em Astroparticle Physics}, vol.~34, no.~9, pp.~679 -- 698,
  2011.

\end{thebibliography}
\bibliographystyle{ieeetr}
\end{document}